\makeatletter
\def\simleq{\mathrel{\mathpalette\gl@align<}}
\def\simgeq{\mathrel{\mathpalette\gl@align>}}
\def\gl@align#1#2{\lower.6ex\vbox{\baselineskip\z@skip\lineskip\z@
     \ialign{$\m@th#1\hfill##\hfil$\crcr#2\crcr\sim\crcr}}}
\makeatother

\documentclass[
    ,final            
  ]
  {aipproc}

\layoutstyle{6x9}

\begin{document}

\title{Meson-Meson and Meson-Baryon Interactions in Lattice QCD}%


\author{Takumi~Doi}{
  address={RIKEN-BNL Research Center, Brookhaven National Laboratory,
  Upton, New York 11973, USA}
}

\author{Toru~T.~Takahashi}{
  address={Yukawa Institute for Theoretical Physics, Kyoto University, 
  Kyoto 606-8502, Japan}
}

\author{Hideo~Suganuma}{
  address={Department of Physics, Kyoto University, 
  Kitashirakawaoiwake, Kyoto 606-8502, Japan}
}

\begin{abstract}
We study the meson-meson
and meson-baryon interactions in lattice QCD.
The simulation is performed on 
$20^3 \times 24$ lattice at $\beta=5.7$
using Wilson gauge action and Wilson fermion at the quenched level.
By adopting one static quark for each hadron as 
``heavy-light meson" and ``heavy-light-light baryon",
we define the distance $r$ of two hadrons and 
extract the inter-hadron potential 
from the energy difference of the two-particle state 
and its asymptotic state.
We find that both of the
meson-meson and meson-baryon potentials are nontrivially weak
for the whole range of $0.2 {\rm fm} \simleq r \simleq 0.8 {\rm fm}$.
The effect of including/excluding the quark-exchange diagrams
is found to be marginal.

\end{abstract}

\maketitle


\section{Introduction}

One of the main goals in hadron physics is
to understand the nature of the nuclear system 
and its interaction from the
fundamental theory.
Although nuclear potential models have been
developed as semi-fundamental theories of nuclei,
the relation between these potential models and
the genuine fundamental theory, QCD, is not
still revealed yet.
In particular, the short-range part of 
the hadron-hadron interaction is of our interest,
because the degrees of freedom of quarks and gluons
in QCD are expected to be most relevant in this range,
while the degrees of freedom
of mesons and baryons are more relevant for the long-range interaction,
as is well established in one-pion exchange
in nuclear force.
Although 
effective short-range interactions such as 
spin-spin
interaction
are proposed so far, it is desirable to analyze from 
nonperturbative first-principle QCD calculations, such as lattice QCD.
It is also worth pointing out that 
recent experimental discoveries of tetra-, penta- and nona-quark
candidates 
require the precise information about the interaction
between meson-meson, meson-baryon and baryon-baryon.
With these motivations,
we study meson-meson,
meson-baryon and baryon-baryon interactions in lattice QCD.
The former two are presented in this paper,
while the latter is presented elsewhere~\cite{nucl_force}.
The analysis in static quark limit is also given
in this proceedings~\cite{okiharu:pot}.

\section{The Formalism and lattice QCD results}

\begin{figure}[bt]
  \includegraphics[width=0.4\textwidth]{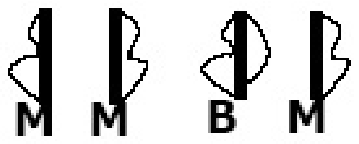}
  \hspace*{8mm}
  \includegraphics[width=0.35\textwidth]{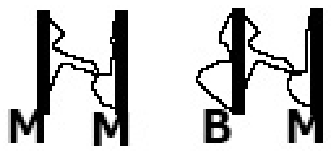}
  \caption{
The diagrams which contribute to
the correlation function. Thick (thin) lines denote 
heavy (light) quarks, and ``M'' (``B'') denotes
meson (baryon).
The right two figures show the light-quark exchange diagrams
and left two show the non-exchange diagrams.
}
\label{fig:diagrams}
\end{figure}

We analyze the meson-meson and meson-baryon interactions
using SU(3)$_c$ lattice QCD.
By calculating the correlation function of hadron-hadron operators,
we evaluate the energy of the hadron-hadron system.
The difference between the energy of the two-particle state
and the sum of the energy of one-particle states
corresponds to the interaction of the particles.
In order to extract the potential as a function of the
distance of two hadrons,
we consider the system with (infinitely) heavy-light mesons
and a heavy-light-light baryon.
In fact, the center of mass of each hadron
is defined by the location of the (heavy) static quark,
and therefore we can define the distance $r$ of 
two hadrons of concern.
This is a more convenient way than performing 
the scattering state analysis using the L{\" u}scher formula,
which is very expensive. The same technique 
is also used in Ref.~\cite{others:M-M}.
We evaluate the correlation function 
$\Pi(\vec{x},\vec{y};t) = 
\langle J(\vec{x},\vec{y};t)\bar{J}(\vec{x},\vec{y};0) \rangle$,
where the operator $J(\vec{x},\vec{y};t)$ is taken as 
$J(\vec{x},\vec{y};t) = J_M(\vec{x};t) J_M(\vec{y};t)$
or
$J(\vec{x},\vec{y};t) = J_B(\vec{x};t) J_M(\vec{y};t)$ 
for a meson-meson or a meson-baryon system, respectively.
Here, the inter-meson/meson-baryon distance 
can be expressed as $r = |\vec{x}-\vec{y}|$.
Using the rotational/translational symmetry, 
we pick up 2-4 spatial configurations 
for $(\vec{x}, \vec{y})$ on the lattice 
and average them at each gauge configuration,
which leads to higher statistics.
In this paper, we consider an
open heavy-flavored pseudoscalar meson 
$J_M (\vec{x};t) \equiv \bar{Q}(\vec{x};t) i \gamma_5 q(\vec{x};t)$, and
a heavy-flavored $\Lambda$ baryon
$J_B (\vec{x};t) \equiv 
\epsilon_{abc}(q_a^T (\vec{x};t)C \gamma_5 q_b(\vec{x};t)) Q_c(\vec{x};t)$, 
where $q$ denotes a light-quark and $Q$ a static quark.
%
Interactions with other meson/baryon systems
are subjects for future work.

In the evaluation of the correlation function,
we have two kinds of diagrams depending
on the flavor assignment for light quarks
in the hadrons.
In fact, if the flavors of light-quarks in two hadrons of concern
are different each other,
only diagrams without quark exchange can
contribute to the correlation function,
as is shown in Fig.~\ref{fig:diagrams}~(left).
On the other hand, if the flavors in two hadrons are identical,
not only non-exchange diagrams but also
exchange diagrams contribute to the correlation function,
as is shown in Fig.~\ref{fig:diagrams}~(left+right).
We analyze both of the two cases,
in order to study the possible effect from 
the Pauli-blocking, direct quark exchange
and/or effective meson exchange.

We perform the Monte Carlo simulations with
the standard Wilson action for $\beta=5.7$ 
($a \simeq 0.19$ fm)
at the quenched level
and generate about 200 gauge configurations.
In order to accommodate two hadrons in the lattice,
we adopt the large lattice of $20^3\times 24$, 
which corresponds to 
$(3.8{\rm fm})^3 \times 4.6{\rm fm}$ 
in the physical scale.
We employ Wilson fermion with the hopping parameter 
$\kappa = 0.1600,0.1625,0.1650$ for a light quark $q$,
which roughly corresponds to $m_\pi \simeq 500-700$ MeV.
%
%
%
%
%

\begin{figure}[tb]
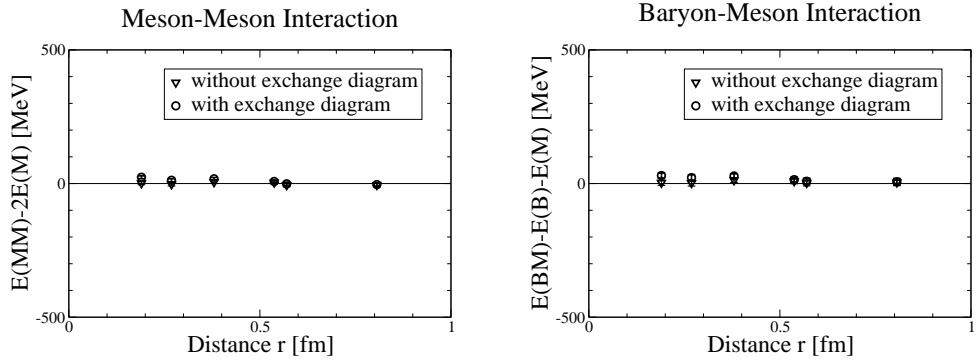

  \includegraphics[width=0.4\textwidth]{exp-meson-meson.1600.proc.eps}
  \hspace*{8mm}
  \includegraphics[width=0.4\textwidth]{exp-baryon-meson.1600.proc.eps}
  \caption{
The meson-meson potential (left) and the meson-baryon potential (right)
in terms of the distance $r$ of two hadrons for $\kappa=0.1600$.
The triangle (circle)  symbols denote the results 
without (with)  quark exchange diagrams.
}
\label{fig:pot}
\end{figure}

In Fig.~\ref{fig:pot}, 
we plot the lattice results of the meson-meson potential 
and the meson-baryon potential for $\kappa=0.1600$.
We find that both of the interaction are rather weak 
for the whole range of distance
$0.2 {\rm fm} \simleq r \simleq 0.8 {\rm fm}$.
%
%
This is quite nontrivial,
considering that the short-range interaction
in nuclear potential at least amounts to
several hundred MeV.
In Fig.~\ref{fig:pot}, 
it is also shown that the 
inclusion/exclusion of quark exchange diagrams
yields marginal effects on the potential.

As a physical reason 
why the potential is so weak,
we examine the possibility that the spin-spin interaction is 
weakened in this lattice QCD simulation, 
because the adopted quark mass is larger 
than the physical light-quark mass.
In fact, in the quark model, the spin-spin interaction is 
suppressed by the (constituent) quark mass.
We, however, find that analyses with lighter quark mass
($\kappa=0.1625, 0.1650$) 
also yield the similar nontrivially weak potential,
which indicates that the above possibility is not a main scenario.
Further studies are in progress, for example, from the 
viewpoint of specific choice of operators, 
to understand the physics of this nontrivial lattice result.



In summary, we have investigated the meson-meson
and meson-baryon interactions in lattice QCD at the quenched level.
By adopting one static quark for each hadron, 
we have defined the distance $r$ of two hadrons and 
have extracted the inter-hadron potentials.
We have found that both of the meson-meson and meson-baryon 
potentials are nontrivially weak. 
The effect of including/excluding the quark-exchange diagrams
has been found to be marginal.

The lattice QCD Monte Carlo simulations have been performed 
on NEC SX-5 at Osaka University and HITACHI SR8000 at KEK.








\bibliographystyle{aipproc}   






\end{document}